\documentclass[graybox, envcountchap]{svmult}

\usepackage{mathptmx}        
\usepackage{amsmath}
\usepackage{amssymb}
\usepackage{color}
\usepackage{helvet}          
\usepackage{courier}         
\usepackage{dirtree}

\usepackage{makeidx}        
\usepackage{graphicx}        
\usepackage{subfig}

\usepackage{multicol}        
\usepackage[bottom]{footmisc}

\usepackage{hyperref}        
\hypersetup{colorlinks=true,urlcolor=blue}

\usepackage[misc]{ifsym}

\newcommand{\hii}{H\,{\sc ii }}
\newcommand{\lsig}{$L(\mathrm{H}\beta) - \sigma$}
\newcommand{\ho}{H$_{0}$}
\newcommand{\oiii}{[O\,{\sc iii}]}
\newcommand{\kmsmpc} {$\rm {km~s^{-1}}~Mpc^{-1}$}

\newcommand{\Msol}{M$_{\odot}$}                           

\def\lsigma{$L-\sigma$~}

\def\hbeta {H$\beta$}
\def\halpha {H$\alpha$}

\newcommand{\kms}{\,km\,s$^{-1}$} 

\makeindex             

\begin{document}


\title{Determination of the local Hubble constant using Giant extragalactic \hii\ regions and \hii\ galaxies}
\titlerunning{Hubble constant via HIIRs and HIIGs} 
\author{David Fernández-Arenas and Ricardo Chávez}
\institute{David Fernández-Arenas (\Letter) \at Canada–France–Hawaii Telescope, 65-1238 Mamalahoa Hwy Kamuela, Hawaii 96743, USA \\ \email{arenas@cfht.hawaii.edu} 
\and Ricardo Chávez (\Letter) \at Instituto de Radioastronom\'ia y Astrof\'isica, UNAM, Campus Morelia, C.P. 58089, Morelia, Mexico \\
Consejo Nacional de Ciencia y Tecnolog\'ia, Av. Insurgentes Sur 1582, 03940 Mexico City, Mexico
\email{r.chavez@irya.unam.mx}}
%

%
\maketitle
\abstract{The relationship between the integrated \hbeta\ line luminosity and the velocity dispersion of the ionized gas of \hii galaxies and giant \hii regions represents an exciting standard candle that presently can be used up to redshifts $z\sim~4$. Locally it is used to obtain precise measurements of the Hubble constant by combining the slope of the relation obtained from nearby ($z<0.2$) \hii galaxies with the zero point determined from giant \hii regions belonging to an `anchor sample' of galaxies for which accurate redshift-independent distance moduli are available (e.g Cepheids, TRGBs). Through this chapter, we present a general description of the method and results obtained so far in the determination of the local value of the Hubble constant and cosmological constraints. We account for the main systematic effects associated with the method and the possibility of improvement in the future. The discussion presented here is the product of an independent approach to the standard methods used in the literature to measure distances and, although at present less precise than the latest SNIa results, it is amenable to substantial improvement.}


\section{Introduction}\label{intro}

The search for standard candles or rulers to measure cosmic distances is one of the fundamental efforts in the last decades in the interface of observational astronomy and cosmology. However, it has been a difficult task for a long time. In this chapter, we present a historical review of the use of \hii galaxies (HIIGs) and giant \hii regions (HIIRs) to carry out this task of measuring distances and its application in the determination of the Hubble constant, from its beginnings to recent years. We discuss its applicability, its potential physical origin and the associated errors which can be improved in the future. 

\begin{svgraybox}
 ``Peculiar" objects from Zwicky list of compact galaxies were identified by \cite{Sargent1970a,Sargent1970b} and further studied by \cite{Sargent1970}. Those that have spectral characteristics indistinguishable from giant extragalactic \hii regions (like 30 Dor in the LMC) were designated by them as ``Isolated Extragalactic \hii Regions". Later on, these compact, dwarf, star-forming galaxies, were dubbed \hii Galaxies \cite{cam86,Terlevich1981,Maza1991}.
\end{svgraybox}  

In the early 1960s, Sersic discovered that the diameters of the largest HIIRs in spiral galaxies increase with the galaxy luminosity.  The use of the HIIRs for the determination of distances goes back to the mid-70s, with the studies of Sandage and Tammann \cite{sandage_and_tamman1974}, mainly with the ``classical" application, an empirical correlations between the diameters of the HIIRs and the parent galaxies luminosity. However, the correlation was not useful beyond 20 Mpc where the angular diameters become comparable with the seeing even for the largest HIIRs.

Later, Melnick in 1977 \cite{Melnick1977} studying the structures of HIIRs,  noticed that their linear diameters also correlated well with the width of their integrated \halpha\ line. This correlation between the velocity dispersion (line-width) in HIIRs and its size is analyzed and established in members of the Local Group and M81 and NGC 2403 group allowing a new determination of the distance to M101. In 1978, Melnick \cite{Melnick1978} explored the close correlation between the linear diameters of HIIRs in late-type galaxies and the width of their global \halpha\ emission and used it to establish a close correlation between this width and the absolute magnitude of the parent galaxies;  this correlation was calibrated in members of the local and NGC 2403 groups, and then used again to estimate the distance to NGC 300 and  M101. The values obtained were consistent with those reported in the literature at that epoch. 


In 1979, the relation between the absolute \hbeta\ fluxes from HIIRs and the corresponding linear diameters and turbulent velocities (width of their integrated \halpha\ line) is examined and the results presented in \cite{Melnick1979}. In this study, a strong correlation was found between the linear diameters and the absolute \hbeta\ luminosities, no longer of the galaxies but of the HIIRs themselves. However, the large scatter precludes its application to distance determination with the properties of the dust being largely responsible for the observed scatter in the relation. 

\begin{svgraybox}
 In 1981, Terlevich and Melnick \cite{Terlevich1981} analyzed the correlations between \hbeta\ luminosities, linear diameters and the widths of the global emission line profiles of 19 giant extragalactic HIIRs in 6 galaxies and 4 HII galaxies. They found that the relations: luminosity$\propto$(linewidth)$^4$ and size$\propto$(linewidth)$^2$ which are valid for pressure-supported stellar systems (elliptical galaxies, bulges of spiral galaxies and globular clusters) are also valid for giant HIIRs. 
\end{svgraybox}  

From this and other considerations, they proposed that giant extragalactic HII regions are self-gravitating systems in which the observed emission-line profile widths represent the velocity dispersion of discrete gas clouds in the gravitational potential of these massive gas—young star complexes.

The analysis also showed that the scatter observed in the \hbeta\ luminosity-linewidth relation for their sample is largely due to a metallicity effect and after correction for this effect, the scatter in the relation is reduced with strong implications as a new method for distance calibrations and although their results are based on a very small sample, they highlight the need for more and better observations of HIIRs. They proposed that these correlations can be calibrated using HIIRs in nearby galaxies for which the distances are known to obtain distances to field spirals and irregular galaxies beyond the Virgo cluster.

From that moment on, the empirical correlations among integrated parameters of HIIRs are considered to have fundamental importance for the understanding of the nature and origin of these objects and represent an independent indicator of distance.

In 1987, with a new homogeneous sample of data for 22  HIIRs in 10 galaxies with known distances, Melnick et. al. \cite{melnick1987} studied the different correlations among integrated parameters of HIIRs. They confirmed the existence of significant correlations between core radius, emission line width, \hbeta luminosity, and oxygen abundance, providing a powerful distance indicator for galaxies with HIIRs. Since, the distance indicator of HIIRs relies on the correlations between \hbeta\ luminosity, the emission line-profile widths and the oxygen abundance of giant HIIRs, they proposed a possible physical scenario where the stellar winds and gravity are the most promising mechanisms for explaining the observations, although not explaining completely the presence of supersonic motions in the ionized gas.  

In 1987, Melnick et. al. \cite{Melnick1987RMxAA} proposed that the most useful application of HIIRs to the determination of distances, is the correlation find  by \cite{Terlevich1981}, which, we will call the \lsig\ relation. Using this extragalactic distance indicator tied to galactic scale via cepheid variables, 41 HIIGs from the group's spectrophotometric catalogue \cite{tel91} and the zero point calibrated with the giant HIIRs in nearby spiral galaxies,  for the first time a value of \ho=$95\pm10$ \kmsmpc was obtained for the Hubble parameter, with the distance calibrator  given by:

\begin{equation}
\log \rm L(\rm H\beta)=\log \frac{\sigma^5}{\rm (O/H)}+41.32
\end{equation}

A year later, Melnick et. al. \cite{melnick1988} calibrated the correlation between luminosity, line width and oxygen abundance. They reported a value for \ho\ of $89.0\pm10$ \kmsmpc, after correcting the flows for Malmquist bias and the radial velocities for the movement of the Local Group. In this study, they discuss the initial potential problems related to the application of the method: the zero point errors and the curvature of the \lsig\ relation. An underestimate of the zero point may arise if many HIIGs are composed of more than one HIIR superimposed along the line of sight. On the other hand, the curvature related to the broadening of the emission lines may be affected by the gravity of the overall galaxy or systems with rotation. If present, this effect will saturate the \lsig\ relation at high luminosities where the line broadening will be dominated by the gravity of the old stellar component. Excluding systems with a linewidth  $>60$\kms\, the correlation between \hbeta\ luminosity, the width of the line and oxygen abundance, HIIGs show the same functional form and the same scatter as that for local HIIRs, suggesting that any broadening of the lines by a source not related to the young stellar population must be small. Additionally, by imposing the condition that the equivalent width of \hbeta$>30$\AA,  effectively the number of galaxies where the light of the young component is severely affected by the underlying older stellar population is reduced.

In the last two decades, with the advance of technology in  new detectors, the new generation of big telescopes and given the potential use of HIIRs and HIIGs as  distance indicators,  a strategy to measure the Hubble constant and to constrain cosmological parameters was devised. It is a long-term program using HIIGs, as  standard candles that can, in principle, be applied out to $z\sim3.5$ \cite{Melnick2000,Plionis2011,terlevich2015} with present-day ground-based instrumentation and up to $z\sim9$ with the Near Infrared Spectrograph (NIRSpec) component of the James Webb Space Telescope (JWST) (\cite{deGraaff2023,Bunker2023} and Chavez et al in preparation). Let's remember that the potential of HIIGs as distance indicators stems from the existence of a correlation between the luminosity of a hydrogen recombination line, e.g. L(\hbeta) (proportional to the number of ionising photons), and the line width of the emission line, $\sigma$.

In what follows, we show the latest advances in the description of the \lsig\ relation, its calibrations with new data sets and the application to the determination of the local value of the Hubble constant.


\section{The L-$\sigma$ relation}

Two of the more recent works related to the \lsig\ relation for HIIGs have been presented in the works of Bordallo \& Telles \cite{Bordallo2011} and Chávez et. al.\cite{Chavez2014} in 2011 and 2014, respectively. They presented new observational datasets of high- and low- spectral resolution. After multi-parametric fits, they found that the \hbeta\ equivalent width is more important for determining 
the scatter in the \lsig\ relation  than the dependence on metallicity, contrary to the suggestion by Terlevich et. al. \cite{Terlevich1981}. Assuming that the EW(\hbeta) is an indicator of the burst's age, the interpretation is that the age of the underlying older stellar population could be contaminating the luminosity of a recent dominant young burst more than its velocity dispersion. Chávez et. al. \cite{Chavez2014} show that the majority of the scatter in the \lsig\ relation is reduced using the size of the burst as a second parameter, which allows us to think of a fundamental plane defined by the luminosity, velocity dispersion and size in HIIGs and HIIRs. 

Both studies conclude that, in addition, part of the scatter in the \lsig\ relation can be reduced by using the objects with the most Gaussian profiles in their emission lines. However, the cost of reducing the samples is of the order of 50\% using this criterion, which can be parameterized in different ways, (see section 4.2 in \cite{Bordallo2011} and sec 6.1 in \cite{Chavez2014}). The main parametric relations presented in the above papers are reproduced in table \ref{tablelsig}. According to different settings, constraints are placed on the slope between 3-5, which could be consistent with a fundamental plane similar to that followed by elliptical galaxies and old star systems supported by gravity. A study on the evolution of these young systems and their scale relationships compared to old systems can be found in Terlevich et. al. \cite{Terlevich2018}. However, the existence of the \lsigma\ relation for HIIRs and HIIGs poses an intriguing question about the origin of the supersonic velocity widths of the emission line profiles. A simple explanation would be that metal rich galaxies produce more massive starbursts, and this should be verified by observations. Another hint about the origin of velocity widths found in HIIGs is provided by the systematic difference found between the width of Balmer lines and \oiii\ lines, the most recent studies relating to ``The \lsigma\ relation for HIIGs in green'' can be found in Melnick et. al. 2017 \cite{Melnick2017}. A possible explanation would be the existence of a well behaved ionization structure, in which more excited ions are concentrated closer to the ionizing sources and densest regions.

\begin{table}
\caption{Fits for the \lsigma\ relation in local HIIGs}
\label{tablelsig}       
\begin{tabular}{p{3.0cm}p{2.0cm}p{2cm}p{2.0cm}p{2.0cm}}
\hline\noalign{\smallskip}
\multicolumn{5}{c}{$\log \rm L = \alpha + \beta \times \log \sigma$} \\
\noalign{\smallskip}\svhline\noalign\\
Linear Regression & Intercept ($\alpha$) & Slope ($\alpha$) & rms & N  \\
\noalign{\smallskip}\svhline\noalign\\
\multicolumn{5}{l}{(BT11, Here the luminosity of \halpha\ has been used instead of \hbeta\ luminosity)} \\
\hline\noalign{\smallskip}

OLS(Y$\mid$X) & 36.21 $\pm$ 0.32 & 3.01 $\pm$ 0.23 & 0.37 & 81$^a$ \\
OLS(X$\mid$Y) & 34.52 $\pm$ 0.38 & 4.18 $\pm$ 0.27&\\
OLS Bisector     & 35.49 $\pm$ 0.32 & 3.51 $\pm$ 0.23&\\
\hline\noalign{\smallskip}

OLS(Y$\mid$X) & 35.29 $\pm$ 0.42&  3.72 $\pm$ 0.31 &0.31&53$^b$\\
OLS(X$\mid$Y) & 33.73 $\pm$ 0.47& 4.85 $\pm$ 0.34& \\
OLS Bisector     & 34.61 $\pm$ 0.41 & 4.22 $\pm$ 0.30&\\
\hline\noalign{\smallskip}

OLS(Y$\mid$X) & 34.80 $\pm$ 0.41 & 4.14 $\pm$ 0.29 &0.29& 37$^c$\\
OLS(X$\mid$Y) & 33.45 $\pm$ 0.53 & 5.13 $\pm$ 0.38& &\\
OLS Bisector    & 34.19 $\pm$ 0.43  & 4.58 $\pm$ 0.30& &\\
\hline\noalign{\smallskip}
\multicolumn{5}{l}{(C14)} \\
\hline\noalign{\smallskip}
Error in both axes & 33.71 ± 0.21 & 4.65 ± 0.14 &0.332& 107$^d$\\
Error in both axes & 33.22 ± 0.27 & 4.97 ± 0.17 &0.332& 69$^e$\\
\noalign{\smallskip}\hline\noalign{\smallskip}
\end{tabular}
$^a$ All galaxies with homogeneous spectrophotometry in. $^b$The sample showing regular Gaussian profiles. $^c$The sample showing regular Gaussian profiles and homogeneous data. $^d$All galaxies with homogeneous data in. $^e$The sample showing regular Gaussian profiles.
\end{table}


\section{The data} 

The use of the \lsigma\ relation as a distance indicator and as a tool to derive the Hubble constant requires accurate determination of both the luminosity and the FWHM or velocity dispersion of the emission lines in giant HIIRs and HIIGs.  Observationally HIIGs and HIIRs represent the youngest Super Star Clusters that can be observed in any detail. The measurements that are required for the calibration of the \lsig\ relation are the following: 

\begin{enumerate}
\item Narrow-band imaging or low-resolution-wide slit spectroscopy  to obtain the integrated \hbeta\ flux.
\item High-resolution spectroscopy to measure the velocity dispersion from the \hbeta\ (or \oiii\ for comparison) line profile.
\end{enumerate}

\begin{figure}
\begin{center}
\includegraphics[scale=0.3]{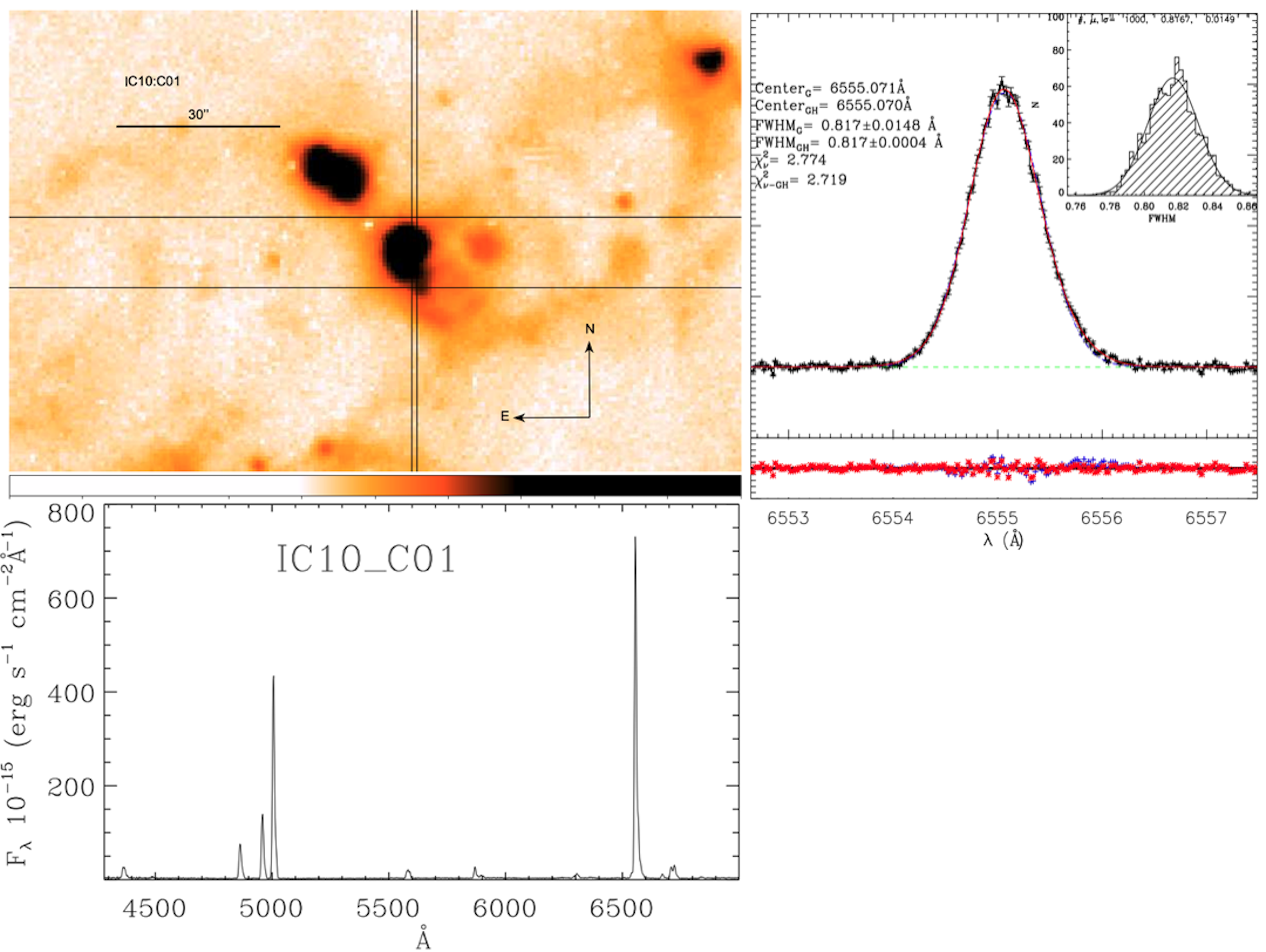}
\caption{\halpha\ image obtained from NASA/IPAC Extragalactic Database (NED), high-resolution profile for the giant HIIR (IC10: 101) and low-resolution spectrum. Taken from \cite{Fernandez2018}}
\label{hiiregions}
\end{center}
\end{figure}

In  figure \ref{hiiregions}, we show as an example the data for one HIIR in a nearby galaxy used in the calibration of the zero-point of the \lsigma\ relation.

To guarantee the best-integrated spectrophotometry and the youngest objects, the following selection criteria are applied:

\begin{itemize}
\item A lower limit for the equivalent width, EW(\hbeta)$>50$\AA (HIIGs and giant HIIRs).
\item An upper limit to the Balmer line widths, $\log\sigma<1.8$ \kms (HIIGs and giant HIIRs).
\item Redshift range $0.01 < z < 0.2$ (HIIGs).
\item Petrosian diameter less than 6 arcsecs (HIIGs).
\end{itemize}

\begin{figure}
\begin{center}
\includegraphics[scale=0.23]{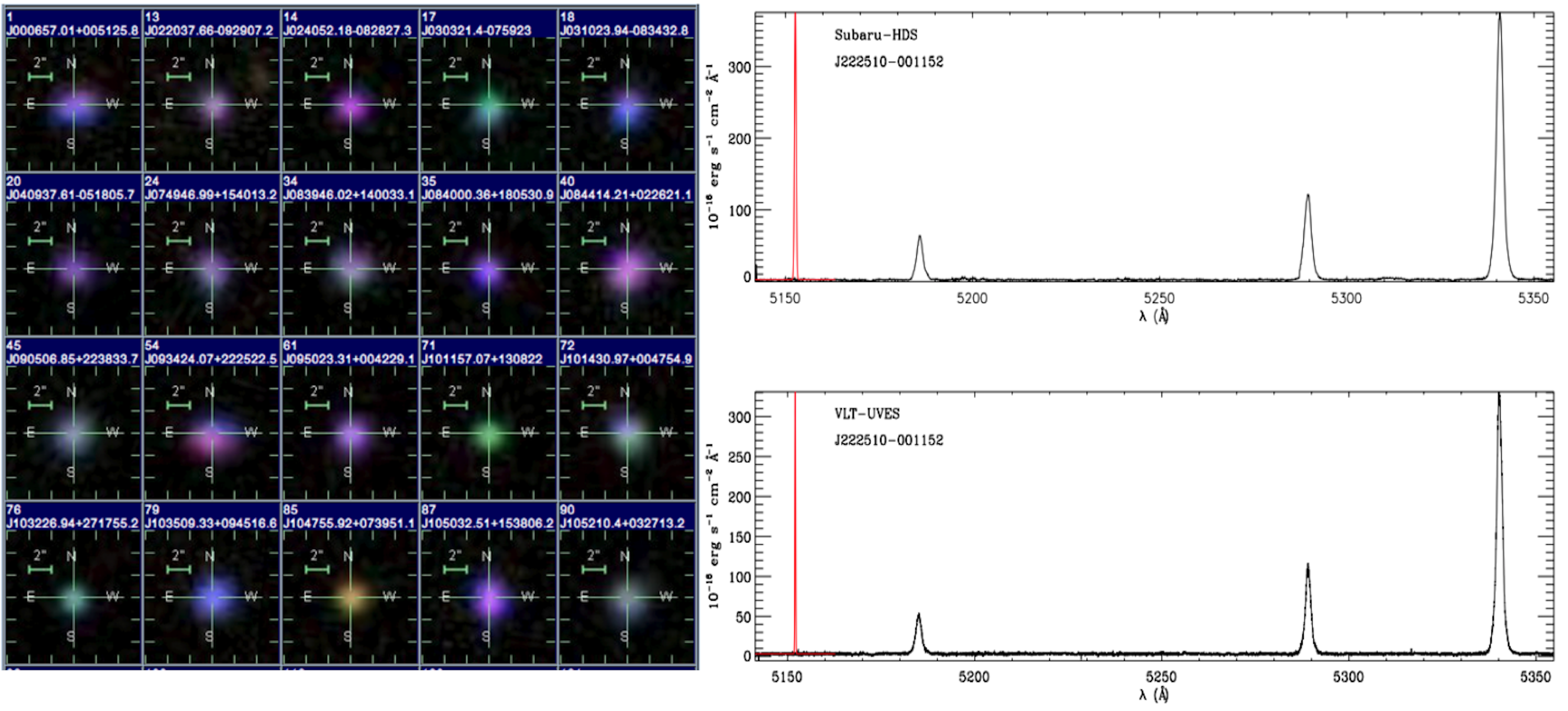}
\caption{A selection of colour images of HIIGs from the  sample of \cite{Chavez2014}. The SDSS name and the index number are indicated in the stamps. The changes in colour are related to the redshift of the object. In the right panel: Examples of the high dispersion spectra obtained for the same object with Subaru HDS (top) and VLT UVES (bottom), showing the region covering \hbeta\ and the \oiii\ lines at $\lambda\lambda$ 4959,5007 \AA. The instrumental profile is shown in red at the left of each spectrum. Taken from \cite{Chavez2014}.}
\label{hiigalaxies}
\end{center}
\end{figure}

Figure \ref{hiigalaxies}  shows typical HIIG in the local universe and their high-resolution spectrum dominated by strong emission lines.

The above criteria allow us to exclude highly evolved regions, to diminish contamination by an underlying older stellar population and to avoid objects with a high rate of escape of ionizing photons. They also  minimize the possibility of including systems supported by rotation or with multiple young ionizing clusters as it was discussed in \cite{melnick1988}. The lower redshift limit in HIIGs is selected to avoid nearby objects that are more affected by local peculiar motions relative to the Hubble flow and the upper limit was set to minimize  cosmological non-linearity effects.

\begin{figure}
\begin{center}
\includegraphics[scale=0.25]{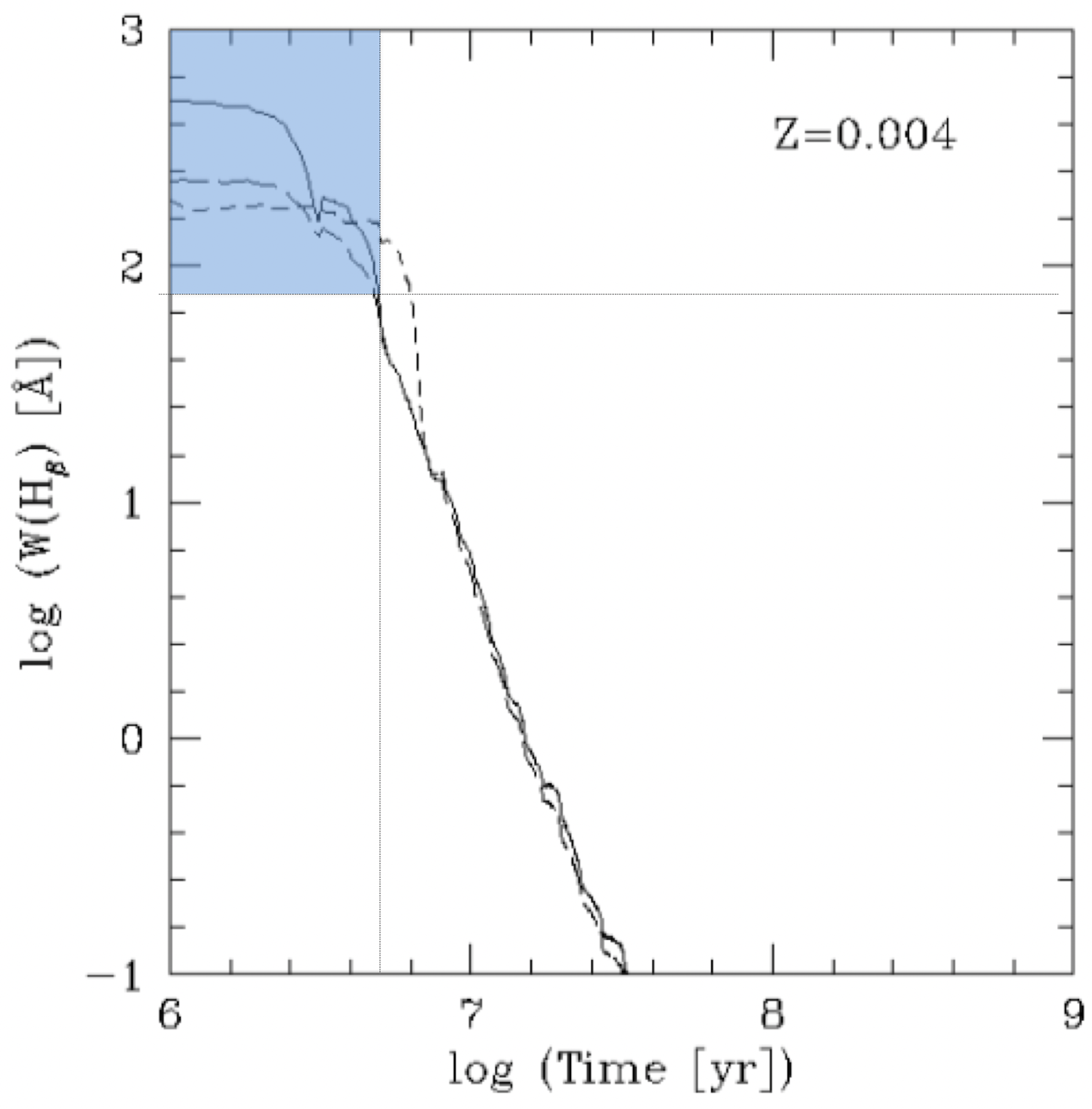}\includegraphics[scale=0.25]{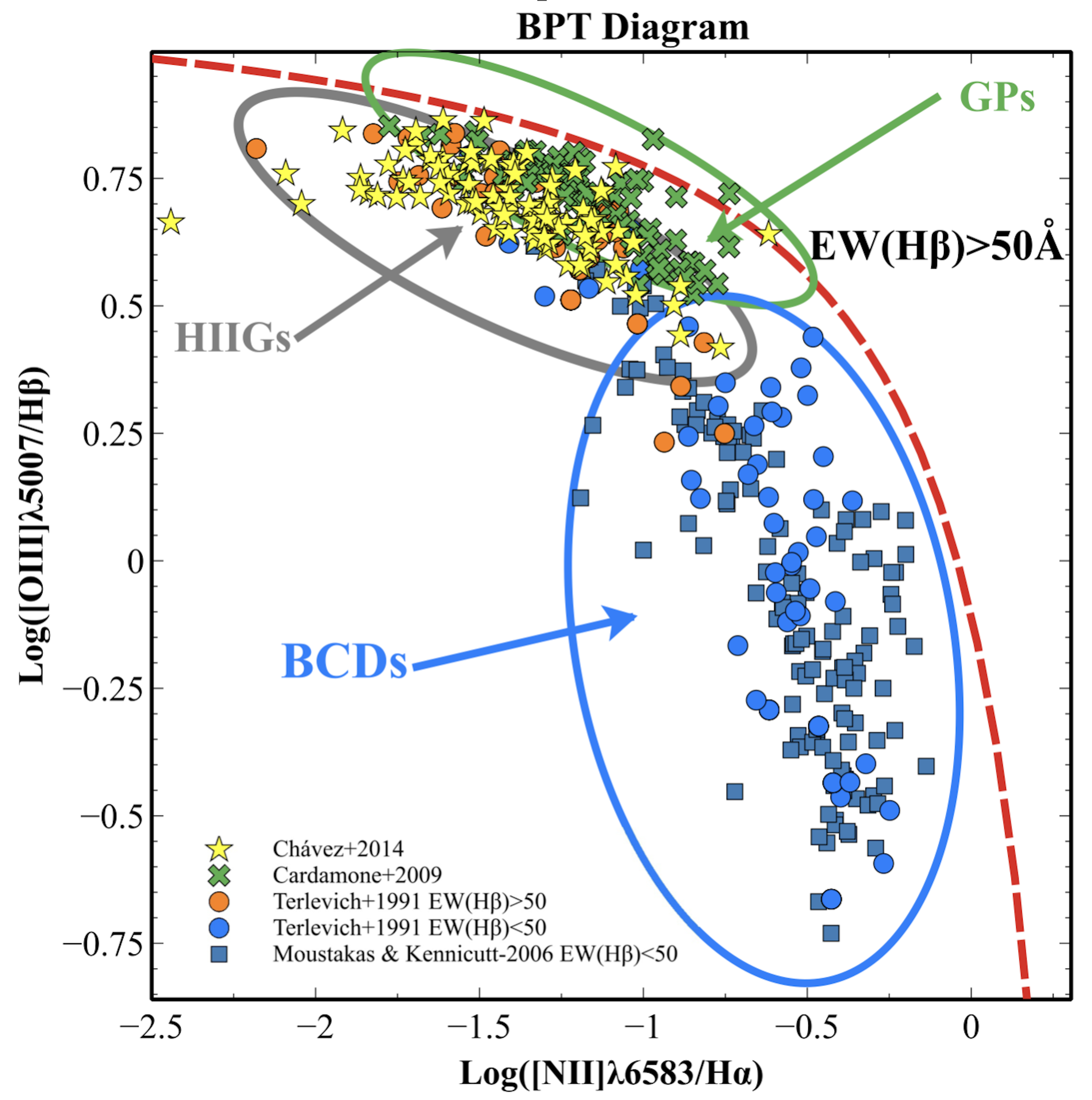}
\caption{The evolution of the \hbeta equivalent width for an instantaneous burst with metallicity $Z= 0.004$ and a Salpeter IMF with an upper limit of 100 \Msol \cite{Leitherer1999}. The blue area marks the \hbeta\ equivalent width of 50\AA,  corresponding to an age of $\sim5$ Myrs.  In the right plot: BPT diagram showing the high excitation level of a sample of HIIGs selected mainly as having high equivalent width in their Balmer emission lines. The solid line represents the upper limit for stellar photoionization, from \cite{Kewley2001}. Taken from  Fernandez PhD Thesis, 2018. }
\label{ewevol}
\end{center}
\end{figure}

In figure \ref{ewevol}, we can see the evolution of the \hbeta\ equivalent width for a simple instantaneous burst with metallicity $Z= 0.004$. In the right panel, we show the location of the HIIGs in the BPT diagrams \cite{Baldwin1981} which are widely used to differentiate the excitation mechanism for star-forming galaxies from AGN, the EW criterion clearly can separate between other similar star-forming ionizing mechanism as Blue Compact Dwarfs and also show  green peas galaxies as a class of HIIGs discovered using different selection criteria. Interesting, in recent years, making evolution studies of these objects the interpretation has been that these young massive stellar clusters, HIIGs and GHIIRs can evolve to form globular clusters and ultracompact dwarf ellipticals in about 12 Gyr so that present-day globular clusters and ultracompact dwarf ellipticals could have originated in conditions similar to those observed in today GHIIR and HIIG \cite{Terlevich2018}. However, this is a topic for another discussion.
 
\subsection{Giant HII Regions and HII galaxies}

HIIG are compact and massive systems experiencing luminous bursts of star formation generated by the formation of young super  clusters (SSCs) with a high  luminosity per unit mass and with properties similar, if not identical, to giant HIIRs. The potential of GHIIRs as distance indicators was originally realized  from the existence of a correlation between the giant HIIR diameter and the luminosity \cite{Sersic1960,Sandage1962} see also \cite{Kennicutt1979}. 

Most HIIG were discovered in objective-prism surveys thanks to their strong narrow emission lines. Currently, in spectroscopic surveys like Sloan Digital Sky Survey (SDSS), they are selected by very large equivalent widths in the Balmer lines. Since the luminosity of HIIG is dominated by the starburst component they can be observed even at large redshifts becoming interesting standard candles. 

So far, we have analyzed a sample of HIIG and GHIIRs containing 217 objects, which can be split into: the anchor sample (36 GHIIRs in 13 local galaxies with distances from primary indicators, \cite{Fernandez2018}), the local sample of HIIG (107 with $z< 0.16$, \cite{Chavez2014}) and a high-$z$ sample based on 29 KMOS, 15 MOSFIRE, 6 XShooter and 24 literature objects (see \cite{Gonzalez2019,Gonzalez2021} for more details). We used these data to constrain cosmological parameters and to determine the local value of the Hubble constant by means of the distance indicator defined by the \lsigma\ relation.

\begin{figure}
\begin{center}
\includegraphics[scale=0.25]{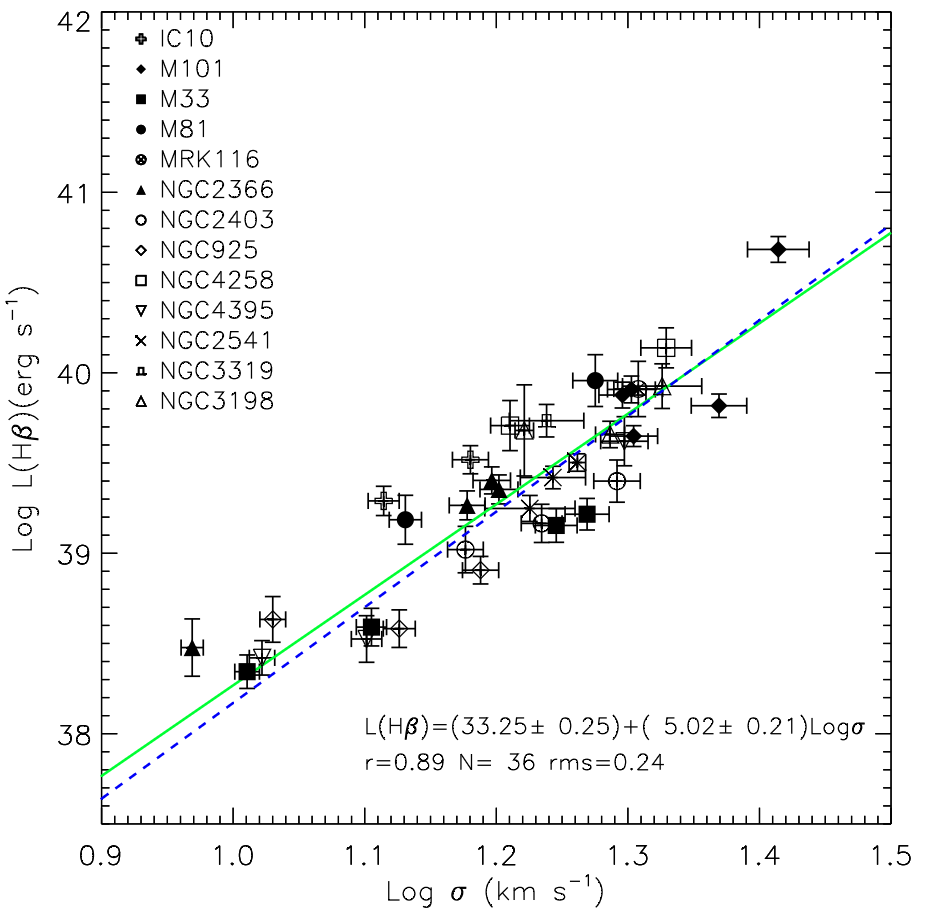}
\caption{\lsigma\ relation for the GHIIRs. The adopted distances to the parent galaxies come from  Cepheids. The green solid line is the fit to the data given in the inset and the dashed line is the fit. Taken from \cite{Fernandez2018}.}
\label{lsigmaGHIIRs}
\label{hubble}
\end{center}
\end{figure}

\section{Methodology}\label{meth}

The Hubble constant is determined as follows: first we fix the slope of the \mbox{\lsigma} relation using the velocity dispersion and luminosity of the HIIGs. The slope is independent of the actual value of \ho. 

To estimate the Hubble constant we use the slope ($\alpha$) of the  \lsigma relation of the HIIG and new GHIIR data (anchor sample) to calibrate the zero point ($Z_p$) of the distance indicator (see figure \ref{lsigmaGHIIRs})  as follows:

\begin{equation}\label{one}
Z_p=\frac{\sum_{i=1}^{36}{ {\it W_i}(\log L_{\rm GHR,i} - \alpha\times \log\sigma_{\rm GHR,i}) }} { \rm \sum_{i=1}^{36}{{\it W_i} } }
\end{equation}
where $L_{GHR,i}$ is the H$\beta$ luminosity of each GHIIR and $\sigma_{\rm GHR,i}$ is the corresponding velocity dispersion. The statistical weights $W_i$ are calculated as:

\begin{equation}\label{zero}
W_i^{-1}= \Bigl(0.4343\frac{\delta L_{\rm GHR,i}}{L_{\rm GHR,i}}\Bigr)^2 + \Bigl(0.4343\alpha\frac{\delta\sigma_{\rm GHR,i}}{\sigma_{\rm GHR,i}}\Bigr)^2 + (\delta\alpha)^2(\sigma_{\rm GHR,i}-<\sigma_{\rm HIIG}>)^2
\end{equation} 
where $<\sigma_{\rm HIIG}>$ is the average velocity dispersion of the HIIG that define the slope of the relation. Thus, the calibrated \lsigma relation or distance estimator is: $ \log L(H\beta) = \alpha \log\sigma+Z_p$. 
To calculate the Hubble constant we minimise the function, 

\begin{equation}\label{mins}
\chi^2(\textrm{H}_{0}  ) = \sum_{i=1}^{N}[{W_i(\mu_i - \mu_{\textrm{H}_{0},i})^2 - ln (W_i) ]}
\end{equation}
where $\mu_i$ is the logarithmic distance  modulus to each HIIG calculated using the distance indicator and the H$\beta$ flux F(H$\beta$) as  

\begin{equation}\label{mods}
\mu_i=2.5[ Z_p + \alpha\times \log\sigma_i - \log F_i(H\beta) - \log 4\pi]
\end{equation}
and $\mu_{\textrm{H}_{0},i}$ is the distance modulus calculated from the redshift using either the linear relation $D_L=zc/\textrm{H}_{0}$ or the full cosmological prescription with $\Omega_{\Lambda}=0.71$.
 
The best value of $\textrm{H}_{0}$ is then obtained minimising $\chi^2$ with statistical weights $W_i^{-1}=\delta\mu_i^2+\delta\mu_{\textrm{H}_{0},i}^2$ calculated as,

\begin{equation}\label{weights}
W_i^{-1}= 6.25 [(\delta Z_p)^2 + \Bigl(0.4343\frac{\delta F_i}{F_i}\Bigr)^2 + \Bigr(0.4343\alpha\frac{\delta\sigma_i}{\sigma_i}\Bigl)^2    +  (\delta \alpha)^2(\sigma_i-<\sigma>)^2]
\end{equation}

\section{Systematics}\label{meth}

Genuine systematic errors are difficult to estimate. However, we have explored and listed a range of parameters to quantify at least part of the systematic error component. In particular, we present different systematic errors that can be taken into account in order to determine the value of \ho\ using giant HIIRs and HIIGs, these effects are described in the following:

\subsection{Age Effects}

Even in our sample of HIIGs and giant HIIRs, chosen to be the youngest systems, it is crucial to verify that the rapid luminosity evolution of the stellar cluster does not introduce a systematic bias in the distance indicator. This would happen, for example, if the average age of the giant HIIRs is different from that of the HIIGs or if luminous and faint HIIGs have different average ages. In general, if velocity dispersion measures mass, younger clusters will be more luminous than older ones for a given velocity dispersion. The evolution effect can be scrambled by the superposition of bursts of different ages along the line of sight. Nevertheless, even small systematics can have a sizable effect on the value of \ho\ so it is important to remove the evolution effect from the data in a similar fashion as for the dust extinction.

The equivalent width of \hbeta\ [EW(\hbeta)]  is a useful age estimator \cite{dottori1981,sta96,mar08} or at least it provides an upper limit of the age of the burst \cite{terlevich2004}. Indeed there is some empirical evidence for this as discussed by  \cite{Melnick2000}  and \cite{Bordallo2011}. \cite{Chavez2014} explored the possibility that the age of the burst is a second parameter, using the EW(\hbeta) as an age estimator in the  \lsig\ relation for HIIGs, and found a rather weak dependence. Different evolution corrections have been explored in recent years, e.g. \cite{Fernandez2018} and  \cite{Telles2018}. This correction must be taken into account in order to improve the estimates of the value of \ho\, but it must still be explored in more detail with better observations especially to characterize the underlying older population.

\subsection{Extinction Effects}

One of the major contributions to the systematic errors is related to extinction correction, in particular the use of different extinction laws. In particular,\cite{calzetti2000} law yields larger values of \ho\  than \cite{Gordon2003} typically by about 1.5$\sim$\kmsmpc. As the \cite{calzetti2000} law was derived from a sample of eight heterogeneous starburst galaxies where only two, Tol 1924-416 and UGCS410  are bonafide  HIIGs and the rest are evolved high metallicity starburst galaxies, we prefer to use the \cite{Gordon2003} extinction curve which corresponds to the LMC2 supershell near the 30 Dorado star-forming region, the prototypical giant HIIR, in the Large Magellanic Cloud. 

\subsection{Metallicity Effects}

In 1981, Terlevich et. al. \cite{Terlevich1981} proposed that oxygen abundance is a good indicator of the long-term evolution of the system. They proposed a simple ‘closed box’ chemical evolution model with many successive cycles of star formation. In it, for each cycle, evolution is traced by the EW(\hbeta) whereas the long-term evolution of the system, spanning two or more cycles, could be traced by the oxygen abundance, which then becomes a plausible second parameter in the \lsig\ correlation.

\subsection{Aperture Effects}

The sensitivity to \hbeta\ photometry could be crucial for the determination of the luminosity especially when  multiple bursts of star-forming regions  and the nearest HIIGs are present. This effect was explored by \cite{Fernandez2018}  by comparing  SDSS and  \cite{Chavez2014} photometric measurements. The results reveal that the values of \ho\ are on average systematically lower by about 4.9$\sim$\kmsmpc for SDSS fluxes and no evolutionary corrections. This systematic difference is reduced to 1.7$\sim$\kmsmpc when evolutionary corrections are included.  The smaller value of \ho\ for the SDSS photometry is related to the systematically steeper slope of the \lsig\ relation compared with that obtained when using  \cite{Chavez2014} photometry, both with and without evolution correction.  Since the lower luminosity HIIGs are also the closer ones, the steepening is a consequence of the smaller SDSS aperture,  compared with that of \cite{Chavez2014}, underestimating the line fluxes of the nearest galaxies. This effect may also introduce a bias in luminosity or velocity dispersion due to the multiplicity seen in some objects (e.g. J084220+115000) as presented by \cite{Fernandez2023}.

\subsection{Redshift Effects}

An upper redshift cutoff of $z=0.1$ was set (S2) instead of $z = 0.16$ (S1) for the HIIGs. Comparing S1 (in which the distances were computed using a flat cosmology with $\Omega_m=0.29$) with sample S2, using the linear Hubble relation for the distances, we found that using $z<0.1$ reduces the value of \ho\ typically by about 2 $\sim$\kmsmpc with a range from 1.1 to 3.3 \kmsmpc. The sensitivity of \ho\ to the actual value of $\Omega_m$ is low, amounting in our case to an uncertainty of about  0.1\% in \ho\ for an uncertainty in $\Omega_m$ of 0.02 \cite{Betoule2014}. This, together with the larger uncertainty when using  S2 drives us to use  S1 with the distance determined using a flat cosmology with  $\Omega_m=0.29$. This reinforces the point that we are not biasing the results by using additional cosmological parameters.

\begin{figure}
\begin{center}
\includegraphics[scale=0.23]{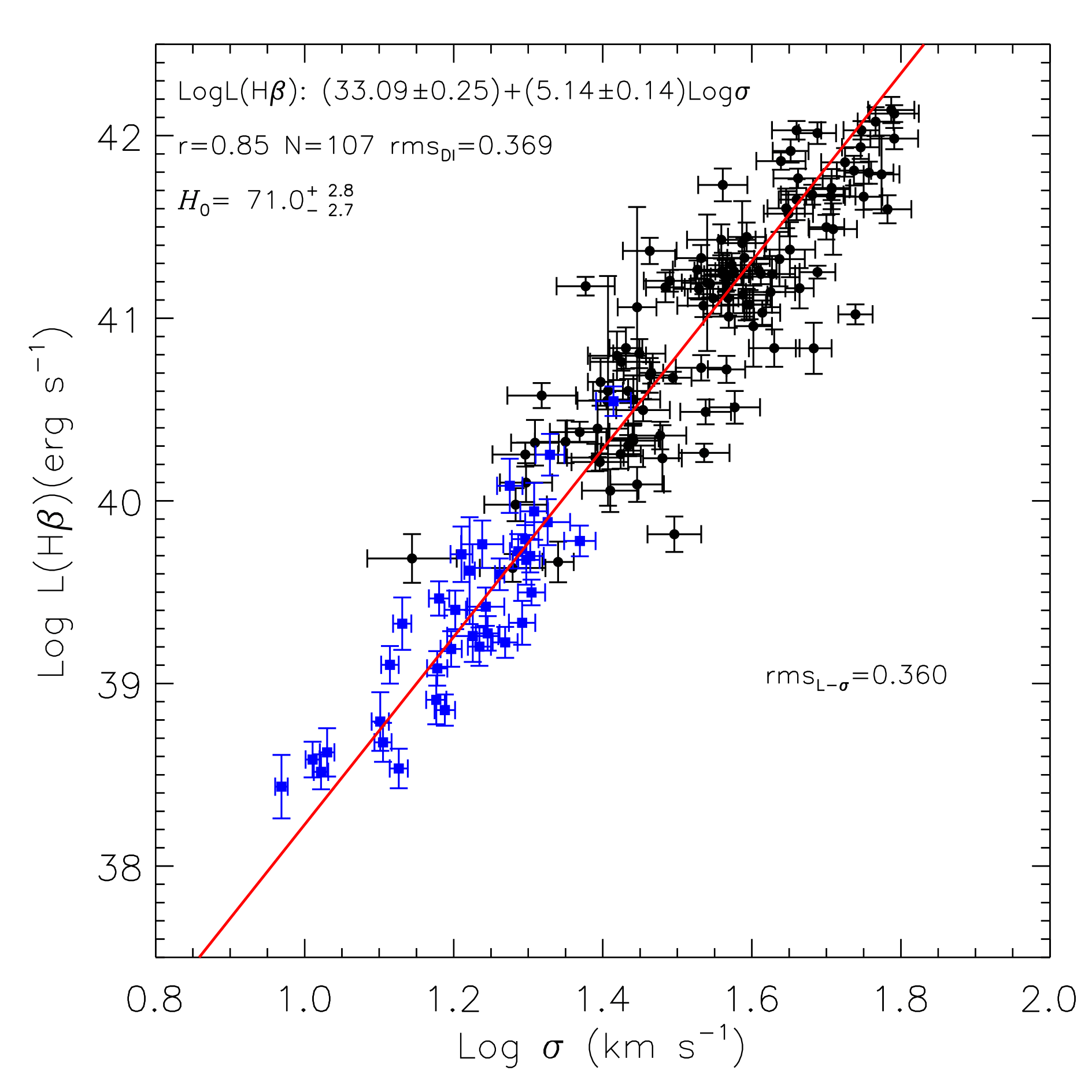}\includegraphics[scale=0.32]{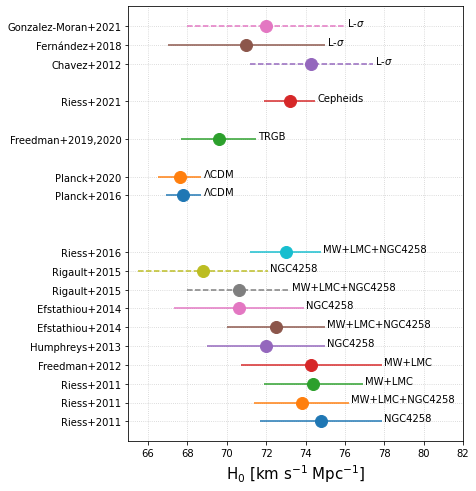}
\caption{Left: The \lsigma relation for the  \protect\cite{Chavez2014} sample using the velocity dispersions in the original paper; the fluxes have been corrected using \protect\cite{Gordon2003} extinction law.  The solid line is the fit to the HIIGs. The inset equation is the distance indicator where the slope is obtained from the fit to the HIIG and the zero point is determined following the procedure described in the text. In the right panel: Our main result incorporating the evolution correction,  is \ho=$71.0\pm3.5$ \kmsmpc (random+systematic) a value that is between the most recent determination from Planck and SNIa. Taken and updated from \cite{Fernandez2018}. We also plot the most recent results derived from the cosmological constrain using HIIGs from Gónzalez-Morán et. al. \cite{Gonzalez2021}.}
\label{hubble}
\end{center}
\end{figure}

\section{Cosmological constraints}
To calculate the parameters of the \lsigma\ relation in a unified way including HIIGs and GHIIRs, we define the following likelihood function:

\begin{equation}
\mathcal{L}\propto \textrm{exp}(-\frac{1}{2}\chi^2_{HII})
\end{equation} where:
 
\begin{equation}
\chi^2_{HII}=\sum_n\frac{(\mu_0(\log f,\log\sigma|\alpha,\beta)-\mu_{\theta}(z|\theta))^2}{\epsilon^2}
\end{equation}
where $\mu_0$ is the distance modulus calculated from a set of observables as:
\begin{equation}
\mu_0=2.5(\alpha+\beta\log\sigma-\log f-40.08)
\end{equation}
$\alpha$ and $\beta$ are the \lsigma\ relation’s intercept and slope, respectively, $\log\sigma$ is the logarithm of the measured velocity dispersion and log f is the logarithm of the measured flux. For HIIG the theoretical distance modulus, $\mu_\theta$, is given as:

\begin{equation}
\mu_{\theta}=5\log d_L(z,\theta)+25
\end{equation}
where $z$ is the redshift, $d_L$ is the luminosity distance in Mpc and $\theta$ is a given set of cosmological parameters. For GHIIRs, the value of $\mu_\theta$ is inferred from primary indicators and finally $\epsilon^2$ are the weights in the likelihood function.

The luminosity distance $d_L$ of the sources tracing the Hubble expansion is employed to calculate the theoretical distance moduli. We define, for convenience, an extra parameter independent of the Hubble constant as:

\begin{equation}
D_L(z,\theta)=(1+z)\int_0^z\frac{dz'}{E(z',\theta)}
\end{equation}
i.e., $d_L=cD_L/\textrm{H}_0$. $E(z.\theta)$ for a flat Universe is given by: 

\begin{equation}
E^2(z,\theta)=\Omega_r(1+z)^4+\Omega_m(1+z)^3+\Omega_w(1+z)^{3y}\textrm{exp}\left(\frac{-3w_az}{1+z}\right)
\end{equation}
with \mbox{$y=(1+w_0+w_a)$}. The parameters $w_0$ and $w_a$ refer to the dark energy equation of state (DE EoS) the general form of which is:

\begin{equation}
p_w=w(z)\rho_w
\end{equation}
with $\rho_w$ the pressure and $\rho_w$ the density of the postulated Dark Energy fluid. Different DE models have been proposed and many are parametrized using a Taylor expansion around the present epoch:

\begin{equation}
w(a)=w_0+w_a(1-a)\Longrightarrow w(z)=w_0+w_a\frac{z}{1+z}
\end{equation}
The cosmological constant is just a special case of DE, given for \mbox{$(w_0, w_a)=(-1, 0)$}, while the so called wCDM models are such that $w_a = 0$ but $w_0$ can take values  $\neq-1$.

\begin{figure}
\begin{center}
\includegraphics[scale=0.28]{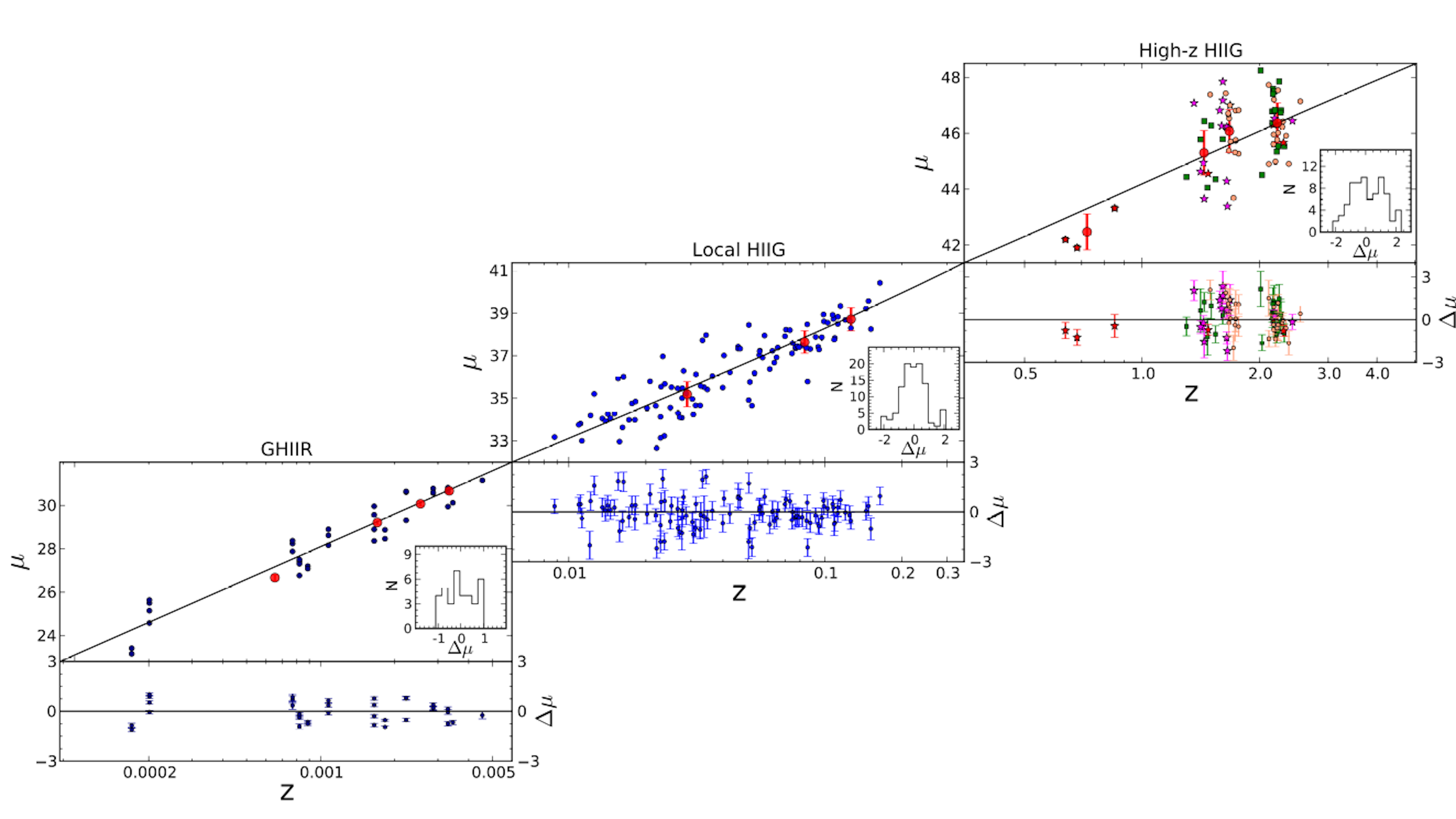}
\caption{Hubble diagram connecting our local and high redshift samples up to $z\sim2.6$. Red circles represent averages of the distance moduli in redshift bins. The continuous line corresponds to \mbox{$\Omega_m=0.249$} and \mbox{$w_0 =-1.18$} (our best cosmological model using only HIIG). The insets show the distribution of the residuals of the fit that are plotted in the bottom panels. Taken from \cite{Gonzalez2021}.}
\label{hubbleDiagram}
\end{center}
\end{figure}

\section{Conclusions}\label{meth}

The accurate determination of the Hubble constant, \ho, is considered one of the most fundamental tasks in the interface between astronomy and cosmology and the \lsig\ relation is a powerful independent distance indicator that has to be explored with more precise data in the future. Here we summarise the main results of the last year in the use of the HIIRs and HIIGs, the Hubble constant and the cosmological constraints.

This chapter presents the determination of the local value of the Hubble constant and observational constraints of the cosmological parameters making use of the GHIIRs in nearby galaxies and HIIG local and at high-$z$, the most recent results are summarized in the Figures \ref{lsigmaGHIIRs},\ref{hubble}, \ref{hubbleDiagram}, where we describe the local \lsig\ relation followed by GHIIRs, the determination of the local value of the Hubble constant, the local \lsig\ relation, the Hubble diagram tracing local GHIIRs up to high-$z$ HIIG,  the space of solutions in the cosmological parameters and the joint analysis of the other tracers of the Hubble expansion. 

From the analysis of different systematic effects, we can infer that they cause changes in the  \lsig\ relation that translate into r.m.s.  variations in the value of \ho\ of 1.5 to 2.5 \kmsmpc.
 
On the other hand, combining HIIGs (the \lsigma\ relation as distance indicator), CMB and BAO yields: \mbox{$\Omega_m = 0.298\pm0.012$} and \mbox{$w_0 = -1.005\pm0.051$}, fully consistent with the $\Lambda$CDM model. It is clear that the solution space of HIIG/CMB/BAO, although less constrained, is certainly compatible with the solution space of SNIa/CMB/BAO.
 

\begin{acknowledgement}
D.F.A  acknowledges the support from the National Science Foundation under grant 2109124 for  SIGNALS: Unveiling Star-Forming Regions in Nearby Galaxies. R.C. gratefully acknowledges support from the CONAHCyT of Mexico under the grant CF2022-320152. D.F.A  and R.C. would like to thank professors Roberto Terlevich and Elena Terlevich for their constant support throughout these years and for their valuable suggestions, which have allowed us to improve the chapter that we present here.
\end{acknowledgement}




\begin{thebibliography}{99}

\bibitem{Sargent1970a} 
W. L. W. Sargent, 
The Astrophysical Journal, \textbf{159}, no. 765 (1970) 
 doi: 10.1086/150353

\bibitem{Sargent1970b} 
W. L. W. Sargent, 
The Astrophysical Journal, \textbf{160}, no. 405 (1970) 
 doi: 10.1086/150443


\bibitem{Sargent1970} 
W. L. W. Sargent et al., 
The Astrophysical Journal, \textbf{162}, no. L155 (1970) 
doi: 10.1086/180644

\bibitem{cam86} 
A. Campbell, et al., 
Monthly Notices of the Royal Astronomical Society, \textbf{223}, no. 811-825 (1986) 
doi: 10.1093/mnras/223.4.811


\bibitem{Terlevich1981} 
R. Terlevich \& J. Melnick, et al., 
Monthly Notices of the Royal Astronomical Society, \textbf{195}, no. 839-851 (1981) 
 doi: 10.1093/mnras/195.4.839

\bibitem{Terlevich1986} 
R. J. Terlevich, 
Star-forming Dwarf Galaxies and Related Objects, \textbf no. 395-402 (1986) 


\bibitem{tel91} 
R. Terlevich, et al., 
Astronomy and Astrophysics Supplement Series, \textbf{91}, no. 285 (1991) 


\bibitem{terlevich2004} 
R. Terlevich, et al., 
Monthly Notices of the Royal Astronomical Society, \textbf{348}, no. 4, 1191-1196 (2004) 
 doi: 10.1111/j.1365-2966.2004.07432.x


\bibitem{terlevich2015} 
R. Terlevich, et al., 
Monthly Notices of the Royal Astronomical Society, \textbf{451}, no. 3, 3001-3010 (2015) 
 doi: 10.1093/mnras/stv1128



\bibitem{Terlevich2018} 
E. Terlevich, et al., 
Monthly Notices of the Royal Astronomical Society, \textbf{481}, no. 268-276 (2018) 
 doi: 10.1093/mnras/sty2325




\bibitem{Kewley2001} 
L. J. Kewley, et al., 
The Astrophysical Journal, \textbf{556}, no. 1, 121-140 (2001) 
 doi: 10.1086/321545

\bibitem{Maza1991} 
J. Maza, et al., 
Astronomy and Astrophysics Supplement Series, \textbf{89}, no. 389 (1991)
 doi:


\bibitem{Sandage1962} 
A. Sandage, 
Problems of Extra-Galactic Research, \textbf{15}, no. 359 (1962) 
 doi: 

\bibitem{sandage_and_tamman1974} 
A. Sandage \& G. A. Tammann, 
The Astrophysical Journal, \textbf{190}, no. 525-538 (1974) 
 doi: 10.1086/152906



\bibitem{Melnick1977} 
J. Melnick, 
The Astrophysical Journal, \textbf{213}, no. 15-17 (1977) 
 doi: 10.1086/155122

\bibitem{Melnick1978} 
J. Melnick, 
Astronomy and Astrophysics, \textbf{70}, no. 157 (1978) 
 doi:

\bibitem{Melnick1979} 
J. Melnick, 
The Astrophysical Journal, \textbf{228}, no. 112-117 (1979) 
 doi: 10.1086/156827
 

\bibitem{melnick1987} 
J. Melnick, et al., 
Monthly Notices of the Royal Astronomical Society, \textbf{226}, no. 849-866 (1987) 
 doi: 10.1093/mnras/226.4.849

\bibitem{Melnick1987RMxAA} 
J. Melnick, et al., 
Revista Mexicana de Astronomia y Astrofisica, vol. 14, \textbf{14}, no. 158-164 (1987) 
 doi:
 
 \bibitem{melnick1988} 
J. Melnick, et al., 
Monthly Notices of the Royal Astronomical Society, \textbf{235}, no. 297-313 (1988) 
 doi: 10.1093/mnras/235.1.297
 

\bibitem{Melnick2000} 
J. Melnick, et al., 
Monthly Notices of the Royal Astronomical Society, \textbf{311}, no. 3, 629-635 (2000) 
 doi: 10.1046/j.1365-8711.2000.03112.x
 
\bibitem{Melnick2017} 
J. Melnick, et al., 
Astronomy and Astrophysics, \textbf{599}, no. A76 (2017) 
 doi: 10.1051/0004-6361/201629728 
 
\bibitem{Plionis2011} 
M. Plionis, et al., 
Monthly Notices of the Royal Astronomical Society, \textbf{416}, no. 4, 2981-2996 (2011) 
 doi: 10.1111/j.1365-2966.2011.19247.x

\bibitem{deGraaff2023} 
de Graaff, et al., 
arXiv e-prints 2023
doi: 10.48550/arXiv.2308.09742

\bibitem{Bunker2023} 
Bunker, et al., 
arXiv e-prints 2023
doi: 10.48550/arXiv.2306.02467

\bibitem{Leitherer1999} 
C. Leitherer, et al., 
The Astrophysical Journal Supplement Series, \textbf{123}, no. 1, 3-40 (1999) 
 doi: 10.1086/313233
 
\bibitem{Bordallo2011} 
V. Bordalo \& E. Telles, et al., 
The Astrophysical Journal, \textbf{735}, no. 1, 52 (2011) 
 doi: 10.1088/0004-637X/735/1/52


\bibitem{Chavez2012} 
R. Chávez, et al., 
Monthly Notices of the Royal Astronomical Society, \textbf{425}, no. 1, L56-L60 (2012) 
 doi: 10.1111/j.1745-3933.2012.01299.x

\bibitem{Chavez2014} 
R. Chávez, et al., 
Monthly Notices of the Royal Astronomical Society, \textbf{442}, no. 4, 3565-3597 (2014) 
 doi: 10.1093/mnras/stu987


 \bibitem{Fernandez2018} 
D. Fernández Arenas, et al., 
Monthly Notices of the Royal Astronomical Society, \textbf{474}, no. 1, 1250-1276 (2018) 
 doi: 10.1093/mnras/stx2710
 
 \bibitem{Fernandez2023} 
D. Fernández Arenas, et al., 
Monthly Notices of the Royal Astronomical Society, \textbf{519}, no. 3, 4221-4240 (2023) 
 doi: 10.1093/mnras/stac3309 
 
 
 
\bibitem{Sersic1960} 
J. L. Sérsic, 
Zeitschrift fur Astrophysik, \textbf{50}, no. 168 (1960) 
 doi:
 

 \bibitem{Kennicutt1979} 
R. C. Kennicutt, 
The Astrophysical Journal, \textbf{228}, no. 394-404 (1979) 
 doi: 10.1086/156858

\bibitem{Telles2018} 
E. Telles \& J. Melnick, et al., 
Astronomy and Astrophysics, \textbf{615}, no. A55 (2018) 
 doi: 10.1051/0004-6361/201732275

\bibitem{Gonzalez2019} 
A. L. González-Morán, et al., 
Monthly Notices of the Royal Astronomical Society, \textbf{487}, no. 4, 4669-4694 (2019) 
 doi: 10.1093/mnras/stz1577

\bibitem{Gonzalez2021} 
A. L. González-Morán, et al., 
Monthly Notices of the Royal Astronomical Society, \textbf{505}, no. 1, 1441-1457 (2021) 
 doi: 10.1093/mnras/stab1385


\bibitem{dottori1981} 
H. A. Dottori \& E. L. D. Bica, et al., 
Astronomy and Astrophysics, \textbf{102}, no. 2, 245-249 (1981) 
 doi:


\bibitem{sta96} 
G. Stasińska \& C. Leitherer, et al., 
The Astrophysical Journal Supplement Series, \textbf{107}, no. 661 (1996) 
 doi: 10.1086/192377


\bibitem{calzetti2000} 
M. L. Martín-Manjón, et al., 
Monthly Notices of the Royal Astronomical Society, \textbf{385}, no. 2, 854-866 (2008) 
 doi: 10.1111/j.1365-2966.2008.12875.x


\bibitem{calzetti2000} 
D. Calzetti, et al., 
The Astrophysical Journal, \textbf{533}, no. 2, 682-695 (2000) 
 doi: 10.1086/308692

\bibitem{Gordon2003} 
K. D. Gordon, et al., 
The Astrophysical Journal, \textbf{594}, no. 1, 279-293 (2003) 
 doi: 10.1086/376774



\bibitem{Betoule2014} 
M. Betoule, et al., 
Astronomy and Astrophysics, \textbf{568}, no. A22 (2014) 
 doi: 10.1051/0004-6361/201423413


\bibitem{mar08} 
M. L. Martín-Manjón, et al., 
Monthly Notices of the Royal Astronomical Society, \textbf{385}, no. 2, 854-866 (2008) 
 doi: 10.1111/j.1365-2966.2008.12875.x


\bibitem{Baldwin1981} 
J. A. Baldwin, et al., 
Publications of the Astronomical Society of the Pacific, \textbf{93}, no. 5-19 (1981) 
 doi: 10.1086/130766



\end{thebibliography}
\end{document}